\documentclass[pre,twocolumn,preprintnumbers,aps,showpacs,amsmath,amssymb]{revtex4}

\usepackage{graphicx}
\usepackage{dcolumn}
\usepackage{bm}

\begin{document}
	
	
	\title{Power law observed in the motion of an asymmetric camphor boat under viscous conditions}
	
	\author{Michiko Shimokawa}
	\author{Masashi Oho}
	\author{Kengo Tokuda}
	\affiliation{Fukuoka institute of Technology, 3-30-1 Wajiro-Higashi, Higashi-ku, Fukuoka 811-0295, Japan}
	
	\author{Hiroyuki Kitahata}
	\affiliation{Department of Physics, Chiba University, Yayoi-cho 1-33, Inage-ku, Chiba 263-8522, Japan }
	\date{\today}
	
	\begin{abstract}
		We investigated the velocity of an asymmetric camphor boat moving on aqueous solutions with glycerol.
		The viscosity was controlled by using several concentrations of glycerol into the solution.
		The velocity decreased with an increase in the glycerol concentration.
		We proposed a phenomenological model, and showed that the velocity decreased with an increase in the viscosity according to power law.
		Our experimental result agreed with the one obtained from our model.
		The results provided an approximation that the characteristic decay length of the camphor concentration profile at the front of the boat was sufficiently shorter than that at the rear of the boat, which was difficult to measure directly.
	\end{abstract}
	
	\maketitle
	
	\section{Introduction}
	We can observe a wide variety of patterns, 
	such as in a traffic jam~\cite{Kikumako, Bando, Sugiyama}, a large-scale ordering of swimming bacteria~\cite{Peng, Nishi},
	a swarm of mosquitoes, a parliament of birds and a school of fish~\cite{Vicsek, Vicsek2, Toner}, formed by living things as self-propelled objects.
	It is one of the challenging studies to understand pattern formations induced by these collective motions.
	
	Similar behaviors also emerge in chemical systems, 
	such as microtubes~\cite{Sumino}, droplets~\cite{Thutupalli, Ohmura, Tanaka}, Janus particles~\cite{Nishi2, janus} 
	and camphor systems~\cite{Suematsu, Suematsu3, Nishimori, Soh, Soh2, Ikura, Suematsu, Kohira, Suematsu2, Nakata, Nagayama, Eric, Kitahata2, Lauga, Suematsu2, Yui, Koyano, Heisler, NishiWakai}.
	Self-propelled objects transform chemical energy into kinetic energy in non-equilibrium systems,
	and move spontaneously as if these were living.
	Recently, a lot of studies have reported  on camphor boats as self-propelled particles in the chemical system~\cite{Suematsu, Nishimori, Kohira, Suematsu2, Nakata, Yui}.
	A camphor boat is made of a plastic sheet attached to a camphor disk.
	When the camphor boat is put on an aqueous surface, the camphor molecules dissolve from the disk under the boat and expand on the surface.
	As the camphor molecules decrease the surface tension of the aqueous phase,
	the camphor boat moves on the aqueous phase spontaneously due to a difference in surface tension around the boat.
	There have been many experimental studies, as well as numerical ones, on the camphor boat.
	Some of the numerical models are based on reaction-diffusion dynamics on the camphor concentration~\cite{Nakata, Nagayama, Eric, Heisler, NishiWakai},
	and the others are based on fluid dynamics~\cite{Soh, Soh2, Lauga}. 
	These models could explain the experimental behaviors in a qualitative manner.
	Basic physical quantities were necessary in order to realize the quantitative correspondence. 
	However, it had been difficult to measure the driving force on the motion of the camphor boat, 
	the surface tension difference between the front and the back of the boat, 
	the diffusion coefficient, 
	the supply rate of camphor molecules from the camphor disk to the water surface, 
	and a relaxation rate before Suematsu {\it et al.} measured these quantitative properties in experiments~\cite{Suematsu2}.
	The results have allowed us to compare the experimental results with theoretical ones quantitatively,
	and have provided a deep understanding of the interesting phenomena of the camphor boat.
	However, they investigated only the situation for pure water as an aqueous phase.
	Thus, we focused on viscosity dependence of the motion with regard to a camphor boat.
	
	As methods to change the viscosities of the aqueous solution under the camphor boat,
	the temperature control of the solution or the use of the solution with different physical concentration is considered.
	We adopted the latter; we used aqueous solutions of glycerol with several glycerol concentration~\cite{Nagayama, Koyano}, 
	and changed the viscosity of the base solution.
	
	In this paper, we investigated the velocity $v$ of the camphor boat for several glycerol concentrations $p$,
	and found that $v$ decreased with an increase in $p$.
	In order to understand the $p$ dependence of $v$, we proposed the mathematical model.
	The model showed a power law $v\sim\mu^{-1/2}$, where $\mu$ is the viscosity of the base solution.
	Our experimental results satisfied the scaling relation obtained from the numerical model.
	The agreement between the experimental result and the theoretical result for the viscosity dependence of $v$ 
	provides an estimation of the concentration field around the camphor boat, which is difficult to measure directly in experiments. 
	
	\section{Experimental procedure}
\begin{figure}[tb]
		\begin{center}
			\includegraphics[width=7cm,clip]{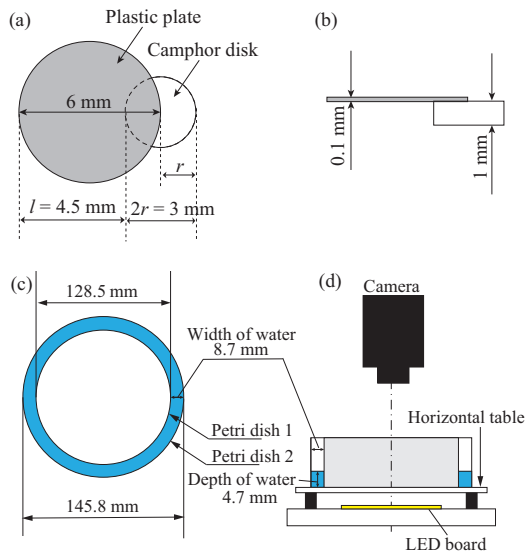}
		\end{center}
		\caption{\label{fig:method}(Color online)
			Schematic drawings of (a) top view and (b) side view of a camphor boat for the measurements of velocities, 
			(c) top view and (d) side view on the annular chamber.}
	\end{figure}

	A round-shape boat as shown in Figs.~\ref{fig:method}(a) and (b) was used to measure the velocity of the camphor boat.
	The boat was composed of a plastic plate (thickness: 0.1 mm) and a camphor disk,
	which was prepared by pressing camphor powder ((+)-Camphor, Wako, Japan) 
	using a pellet die set for the preparation of samples on Fourier transform-infrared (FT-IR) spectroscopy analysis.
	The diameter and the thickness of the camphor disk were 3.0 mm and 1.0 mm, respectively.
	The plastic plate was cut in a circle with a diameter of 6.0 mm, 
	and the camphor disk was attached to the edge of the flat circular plastic plate using an adhesive (Bath bond Q, KONISHI, Japan), 
	so that a half of the camphor disk was outside of the plastic sheet. 
	This round-shape camphor boat moved toward the direction of the plastic sheet.
	
	An annular glass chamber was used,
	which was composed of two petri dishes with different diameters as shown in Figs.~\ref{fig:method}(c) and (d).
	Inner and outer diameters were 128.5 mm and 145.8 mm, 
	and the channel width of the chamber was thus 8.7 mm.
	As it is known that the velocity is sensitive to the depth of water~\cite{Yui}, 
	the chamber was put on the clear horizontal plate.
	The solution was poured into the chamber so that the depth of the solution was 4.7 mm, 
	which was glycerol (Glycerol, Wako, Japan) and water mixed at several mass ratios $p$,
	i.e. $p$ is a percentage of a glycerol mass in the mixed solution.
	We investigated physical properties of the solution, such as the viscosity, the surface tension, and 
	the camphor solubility against glycerol concentration $p$.
	The detailed results are shown in Appendix A.
	The camphor boat was put on the surface of the solution in the glass chamber, 
	and then it started to move spontaneously.
	For a visualization of the motion, a LED board was placed under the horizontal plate.
	The motion of the boat was captured with a digital video camera (HDR-FX1, SONY, Japan) from the top of the chamber.
	Obtained movies were analyzed using an image-processing system (ImageJ, Nature Institutes of Health, USA). 
	
	\section{Experimental Results}
\begin{figure*}[tb]
		\begin{center}
			\includegraphics[width=14cm,clip]{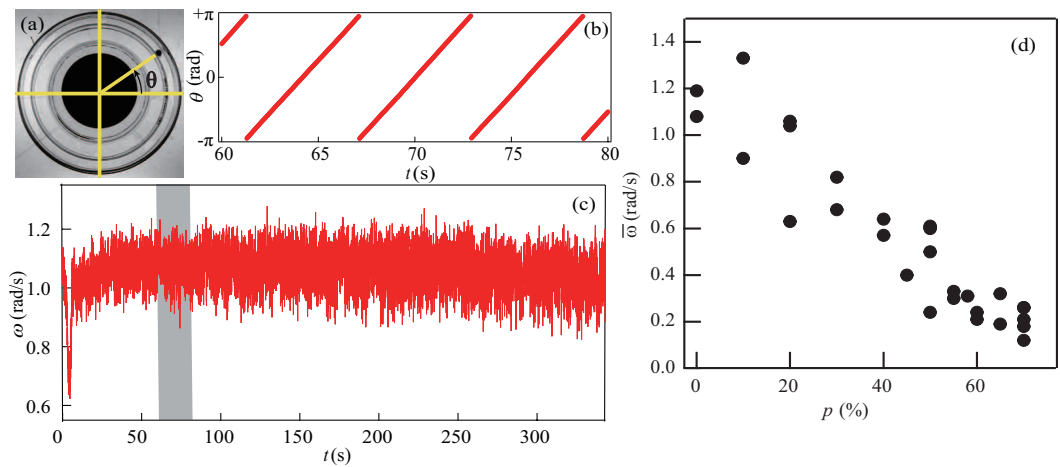}
		\end{center}
		\caption{\label{fig:velo}(Color online)
			(a) Snapshot of the camphor boat motion. 
			(b) Time series of the position $\theta$ of a camphor boat moving on water ($p=0$), 
			where $\theta$ is the angle shown in Fig.~\ref{fig:velo}(a). 
			(c) Time series of angular velocity $\omega$ of the camphor boat, where $\omega=\Delta\theta/\Delta t$ for each frame.
			The gray region corresponds to the time range shown in Fig.~\ref{fig:velo}(b).
			(d) Dependence of $\overline{\omega}$ on $p$, where $p$ is the glycerol concentration 
			and $\overline{\omega}$ is the angular velocity obtained from linear fitting of time series as shown in Fig.~\ref{fig:velo}(b).}
	\end{figure*}

	We investigated the velocity of the camphor boat on the solutions of various glycerol concentration $p$.
	The position of the camphor boat is described as a radial angle $\theta$ in the annular chamber, as shown in  Fig.~\ref{fig:velo}(a).
	Analyses of the videos captured by the digital video camera provide the position $\theta$ at time $t$,
	where $t=0$ corresponds to the time when the boat finished three laps along the chamber after the boat had been put on the surface of the solution.
	In Fig.~\ref{fig:velo}(b), $\theta$ had a constant gradient in time, 
	that is to say, the camphor boat moved with a constant velocity. 
	Figure~\ref{fig:velo}(c) shows a time series of the angular velocity $\omega = \Delta\theta/\Delta t$,
	where $\Delta t =1/30$ s for one frame of the video camera and $\Delta\theta$ is an angular difference between $t$ and $t+\Delta t$.
	In Fig.~\ref{fig:velo}(b), the expanded plot is shown for the time region corresponding to the gray region in Fig.~\ref{fig:velo}(c).
	The angular velocity $\omega$ in the region fluctuated around the average value 1.08 rad/s.
	The similar tendency was observed at 50 s $\lesssim t \lesssim$ 200 s, 
	i.e. $\omega$ increased with time and had noisy data before $t\sim10$ s, 
	and $\omega$ began to decrease after $t\sim250$ s.
	Therefore, we investigated $\omega$ at 60 s $\lesssim t \lesssim$ 180 s, 
	during which $\omega$ had almost a constant value for time.
	Next, we investigated the angular velocity for $p$ as shown in Fig.~\ref{fig:velo}(d).
	The vertical and horizontal axes in Fig.~\ref{fig:velo}(d) show the angular velocity $\overline{\omega}$ and concentration $p$.
	The $\overline{\omega}$ was obtained from the linear fitting of time series as shown in Fig.~\ref{fig:velo}(b).
	The values of the errors for each $\overline{\omega}$ were lower than $10^{-3}$ rad/s.
	As shown in Fig.~\ref{fig:velo}(d), $\overline{\omega}$ decreased with an increase in $p$.
	
	\section{Mathematical Model \label{sec:model}}
	The glycerol concentration $p$ of the solution was controlled in our experiments, which led to a change in the viscosity $\mu$ shown in Appendix A.
	In this section, we consider a viscosity dependence of the camphor boat velocity.
	Now, the annular glass chamber used in our experiments is recognized as a one-dimensional channel with an infinite length. 
	
\begin{figure}[tb]
		\begin{center}
			\includegraphics[width=7cm,clip]{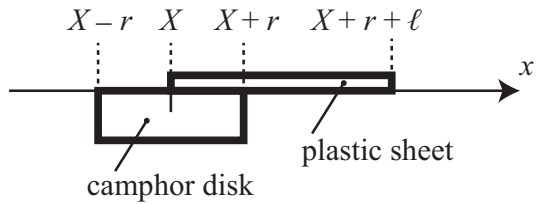}
		\end{center}
		\caption{\label{fig:model}
			Illustration of  side view of a camphor boat.}
	\end{figure}

	The time evolution equation of the camphor boat in a one-dimensional system (The spatial coordinate is represented as $x$) is given as
	\begin{align}
		m\frac{d^2X}{dt^2} = -h\frac{dX}{dt}+F,
		\label{eq:motion}
	\end{align}
	where $m$, $X$, $h$ and $F$ are the mass, 
	the center of mass,
	the friction coefficient of the camphor boat,
	and the driving force exerted on the moving camphor boat, respectively.  
	We assume that $h$ is proportional to viscosity $\mu$ such as $h=K\mu$, where $K$ is a constant ($K > 0$).
	The assumption has been used in many previous papers \cite{Eric,Koyano,Nagayama,Nakata,Nishimori,Suematsu,Suematsu2,Kohira,Soh,NishiWakai,Heisler},
	and it was also reported that the viscous drag on the mobility of thin film in Newtonian fluid obeyed a linear relationship with the fluid viscosity \cite{stone}. 
	Therefore, we considered that the assumption $h=K\mu$ is natural \cite{viscous drag}.
	The driving force $F$ is described as
	\begin{align}
		F = w[\gamma(c(X+r+\ell))-\gamma(c(X-r))],
		\label{eq:driving}
	\end{align}
	where $w$ is the width of the camphor disk.
	Here, we consider that the positions of the front and the back of the boat are shown as $x=X+r+\ell$ and $x=X-r$, 
	where $r$ and $\ell$ are the radius of the disk and the size of the boat as defined in Fig.~\ref{fig:model}.
	The surface tension $\gamma$ depends on the concentration $c$ of the camphor molecules at the surface of the solution,
	and we assume the linear relation as 
	\begin{align}
		\gamma=\gamma_0-\Gamma c,
		\label{eq:surface}
	\end{align}
	where $\gamma_0$ is the surface tension of the base solution without camphor and $\Gamma$ is a positive constant.
	
	The time evolution on the camphor concentration $c$ is shown as
	\begin{align}
		\frac{\partial c}{\partial t} = D \frac{\partial^2 c}{\partial x^2}-ac+f(x-X),
		\label{eq:concentration}
	\end{align}
	where $a$ is the sum of sublimation rate and dissolution rate of the camphor molecules on solution surface, $D$ is the diffusion coefficient of the camphor molecule, and $f$ denotes the dissolution rate of the camphor molecules from the camphor disk to the aqueous solution surface.
	As for the term $f(z)$, we apply the following description,
	\begin{align}
		f(z) = \begin{cases}
			f_0, & ({-r < z <r}),\\
			0, & \text({\rm otherwise}).
		\end{cases}
		\label{eq:provide2}
	\end{align}
	That is to say, the dissolution of camphor molecules from the disk occurs at $-r < z < r$.
	The above equation does not include Marangoni effect directly, 
	although the flow has an influence on the camphor concentration.
	The previous paper \cite{Kitahata2} showed that Eq. (\ref{eq:concentration}) was reasonable 
	if $D$ was recognized as the spatially uniform effective diffusion coefficient of the camphor to include the transportation by the flow.
	In addition, this spatially-uniform effective diffusion coefficient is supported by the experimental results that the diffusion length is proportional to the square root of elapsed time \cite{Suematsu}.
	
	\section{Theoretical analysis}
	Our experimental results showed that the camphor boat moved with a constant velocity in time as shown in Fig.~\ref{fig:velo}.
	Thus, we should consider solutions for the motion of the camphor boat with a constant velocity $v$ in $x$-direction,
	i.e. $X = vt$.
	From this condition, Eq. (\ref{eq:motion}) leads to
	\begin{align}
		-hv+F=0.
		\label{eq:motion1}
	\end{align}
	By setting $\xi=x-vt$ and $c=c(\xi)$, Eq. (\ref{eq:concentration}) provides 
	\begin{align}
		-v\frac{dc}{d\xi} = D \frac{d^2 c}{d\xi^2}-ac+f(\xi).
		\label{eq:concentration1}
	\end{align}
	Equation (\ref{eq:concentration1}) leads to the following solutions
	\begin{align}
		c(\xi) = \begin{cases}
			\beta_1 \exp \big(\lambda_-(\xi-r)\big), & ({\xi > r}),\\
			\dfrac{f_0}{a}+\alpha_2 \exp\big(\lambda_+\xi)+\beta_2 
			\exp\big(\lambda_-\xi\big), & ({-r <\xi < r}),\\
			\alpha_3 \exp \big(\lambda_+(\xi+r)\big), & ({\xi < -r}),
		\end{cases}
		\label{eq:provide}
	\end{align}
	where
	\begin{align}
		\lambda_\pm= -\frac{v}{2D}\pm\frac{\sqrt{v^2+4Da}}{2D},
		\label{eq:lambda_pm}
	\end{align}
	\begin{align}
		\beta_1 = \frac{f_0\lambda_+}{a(\lambda_+-\lambda_-)}(1-\exp(2\lambda_-r)), 
		\label{eq:beta1}
	\end{align} 
	\begin{align}
		\alpha_2= \frac{f_0\lambda_-\exp(-\lambda_+r)}{a(\lambda_+-\lambda_-)},
		\label{eq:alpha2}
	\end{align}
	\begin{align}
		\beta_2 = \frac{f_0\lambda_+\exp(-\lambda_-r)}{a(\lambda_+-\lambda_-)}, 
		\label{eq:beta2}
	\end{align}
	\begin{align}
		\alpha_3 = -\frac{f_0\lambda_-}{a(\lambda_+-\lambda_-)}(1-\exp(-2\lambda_+r)). 
		\label{eq:alpha1}
	\end{align}
	
	Equations (\ref{eq:provide})-(\ref{eq:alpha1}) provide
	\begin{align}
		F =& -\Gamma w \left[\beta_1 \exp \left(\lambda_-\ell \right)-\alpha_3  
		\right]
		\nonumber \\
		=&-\frac{\Gamma w f_0}{a \left(\lambda_+-\lambda_-\right)}  
		\left[\lambda_+ \left(1-\exp \left(2\lambda_-r \right) \right) \exp  
		\left(\lambda_-\ell \right) \right. \nonumber\\
		& \left. + \lambda_- \left(1-\exp \left(-2\lambda_+r\right) \right) \right].
	\end{align}
	
	As $v$ is sufficiently large in our experiments, we assume $r\ll1/\lambda_+$ and $\ell\gg1/\left|\lambda_-\right|$.
	Then, $\lambda_+\sim a/v$ and $\lambda_-\sim -v/D$, which lead to
	\begin{align}
		F = & -\frac{\Gamma w f_0}{a(v/D)} \left[\frac{a}{v} \left(1-\exp  
		\left(-\frac{2vr}{D} \right)\right)\exp\left(-\frac{v}{D}\ell\right)  
		\right. \nonumber \\ &
		\left.-\frac{v}{D}
		\left(1-\exp\left(-\frac{2ar}{v}\right)\right)\right]
		\nonumber\\
		\simeq & -\frac{\Gamma wf_0D}{av}\left(-\frac{v}{D}  
		\right)\left(\frac{2ar}{v} \right)
		\nonumber\\
		= & \frac{2\Gamma wf_0r}{v}.
		\label{eq:F}
	\end{align}
	As $F = K\mu v$ from Eq.~(\ref{eq:motion1}), 
	\begin{align}
		K\mu v = \frac{2\Gamma w f_0 r}{v}.
		\label{eq:F1}
	\end{align}  
	From Eq.~(\ref{eq:F1}), we obtain
	\begin{align}
		v=\sqrt{\frac{2\Gamma wf_0r}{K\mu}}.
		\label{eq:v}
	\end{align} 
	
	Equation~(\ref{eq:v}) shows a power law $v\propto\mu^{-1/2}$, if other parameters such as $\Gamma, w$ and $f_0$ are independent of $\mu$.
	The power law with the index $-1/2$ is an interesting result, 
	since Stokes relation naturally suggests another relation; $v \propto \mu^{-1}$~\cite{fluid}.
	
	\section{Numerical results}
	
	In the theoretical analysis, we have assumed the solution depending on $\xi = x - vt$. However, the supposed mathematical model has other symmetries and whether the considered solution depending on $\xi$ is an attractor or not should be checked. Therefore, we performed numerical calculations based on equations in Sec.~\ref{sec:model}. For numerical calculation, we considered a one-dimensional array with a spatial step of $\Delta x = 0.1$. The spatial size of the considered system was 1000 with periodic boundary condition, and we adopted Euler method with time step $\Delta t = 10^{-3}.$ As for the spatial derivative, we used explicit method. The parameters are set to be $m = 0.1$, $w = 1$, $\Gamma = 1$, $r=1$, $\ell = 1$, $D = 1$, $a = 1$, and $f_0 = 1$. In the discretization process, the first-order interpolation was adopted for Eqs.~\eqref{eq:driving} and \eqref{eq:provide2}. The parameter $h$ corresponding to the viscosity $\mu$ was changed, and we investigated the time development of the camphor boat position and camphor concentration profile.
	
	\begin{figure}[tb]
		\begin{center}
			\includegraphics[width=7cm,clip]{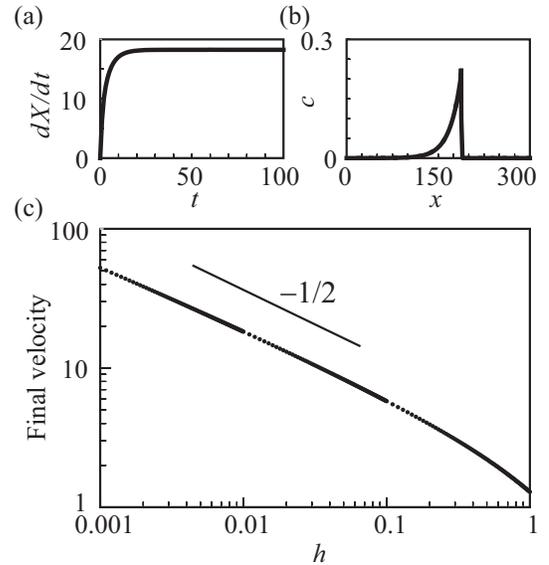}
		\end{center}
		\caption{\label{fig:sim}Numerical results. (a) Time course of camphor boat velocity $dX/dt$ for $h=0.01$. (b) Camphor concentration profile $c(x)$ for $h=0.01$ at $t=1000$, when the camphor boat velocity reached a constant value. The position of the camphor boat was $X \simeq 188.6$. (c) Final velocity ($t = 1000$) depending on $h$, which is proportional to viscosity. The power law $v \propto h^{-1/2}$ holds for smaller $h$.}
	\end{figure}
	
	In Fig.~\ref{fig:sim}, the numerical results are shown. In Fig.~\ref{fig:sim}(a), the time development of camphor boat velocity is shown. The camphor boat velocity is saturated to a constant value. The camphor concentration profile after the velocity became constant ($t = 1000$) is shown in Fig.~\ref{fig:sim}(b). The camphor concentration profile was asymmetric with regard to the camphor boat position $x = X \simeq 188.6$. After reaching a constant velocity, the concentration profile did not change the shape but shifted in a positive $x$-direction. Thus, we can guess that the solution with regards to $\xi = x - vt$ is an attractor of this system. We have also confirmed that the solution converged to this attractor from other initial conditions (data not shown). The mathematical analysis on this convergence to the solution depending on $\xi$ remains and it may be possible to approach such mathematical problem by considering Lie group symmetry \cite{Olver}.
	
	The final velocity against $h$ is shown in Fig.~\ref{fig:sim}(c). For the regime of $h$ smaller than 0.1, the power law $v \propto h^{-1/2}$ held, where $h$ is proportional to the viscosity $\mu$ in the present framework. In the theoretical analysis, we assumed  $r \ll 1/\lambda_+$ and $\ell \gg 1/\left|\lambda_- \right|$, which is equivalent to $aD/v^2 \ll 1$, as will be discussed in detail in the following section. Since the final velocity is nearly equal to 5 for $h \sim 0.1$, and $a = D = 1$, the divergence from the power law originates from the breakdown of the assumption in the analysis.
	
	\section{Discussion}
	Our model showed a power law $v\sim\mu^{-1/2}$ under the assumptions that $r\ll1/\lambda_+$ and $\ell\gg1/\left|\lambda_-\right|$.
	In this section, we compare experimental results with the numerical results in Eq.~(\ref{eq:v}) in order to check whether our model is reasonable.
	Equation~(\ref{eq:v}) has several parameters such as $\Gamma$, $w$, $f_0$, $r$, $K$, and $\mu$.
	Since similar camphor boats were used, $w$, $r$, and $K$ were constant values in our experiments.
	We investigated the dependence of the other parameters, i.e., $\Gamma$, $f_0$, and $\mu$, on the glycerol concentration $p$ in Appendix A.
	Equation~(\ref{eq:surface}) showed $\Gamma=(\gamma_0 - \gamma)/c$.
	As $(\gamma_0 - \gamma)$ was independent of $p$ in our measurements,
	we considered that $\Gamma$ was constant.
	The supply rate $f_0$ corresponds to $\Delta M$, 
	which is a loss of a camphor disk per unit time in our experiments,
	and we found that $\Delta M$ decreased with an increase in $p$.
	The viscosity $\mu$ of the base solution increased with $p$.
	Thus, $f_0$ and $\mu$ in Eq.~(\ref{eq:v}) are functions of $p$. 
	In addition, the angular velocity is proportional to the camphor boat velocity in our experiments.
	
\begin{figure}[tb]
		\begin{center}
			\includegraphics[width=7cm,clip]{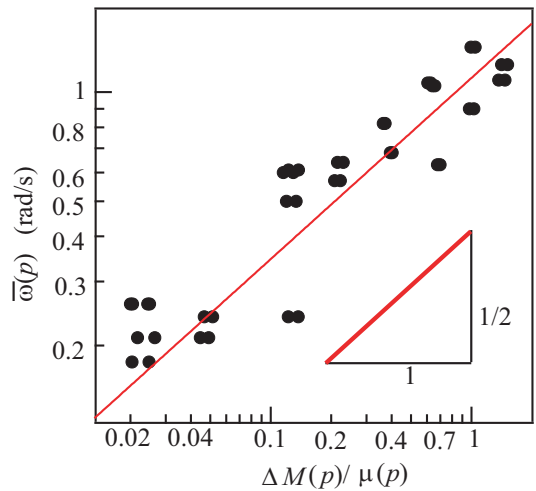}
		\end{center}
		\caption{\label{power}(Color online)
			Relationship between $\Delta M/\mu$ and $\overline{\omega}$, 
			where $\Delta M$, $\mu$, and $\overline{\omega}$ are a weight loss of a camphor disk per one second,
			the viscosity of the base solution, and the angular velocity of the camphor boat, respectively. 
			The solid line shows the numerical result; $\overline{\omega}\sim\sqrt{\Delta M/\mu}$ in Eq.~\eqref{eq:v2}.}
	\end{figure}

	From the above discussion, Eq.~\eqref{eq:v} leads to
	\begin{align}
		\overline{\omega}(p) \propto\sqrt{\frac{\Delta M(p)}{\mu(p)}}.
		\label{eq:v2}
	\end{align} 
	Figure~\ref{power} shows a relationship between $\Delta M/\mu$ and $\overline{\omega}$ obtained from our experiments.
	The result almost agrees with the solid line in Eq.~(\ref{eq:v2})~\cite{Delta_M}.
	
	The power law was obtained under the assumptions that $r\ll1/\lambda_+$ and $\ell\gg1/\left|\lambda_-\right|$,
	which is equivalent to $aD/v^2\ll1$.
	Since $\sqrt{D/a}$ corresponds to a characteristic decay length of the camphor concentration profile, 
	and $v/a$ is a distance of the camphor boat motion during the characteristic time during which the concentration field keeps the memory, 
	the assumption means that the characteristic length for the camphor concentration profile is sufficiently smaller than the characteristic length for the camphor boat motion. 
	In such a case, the camphor concentration profile should be asymmetric with respect to the camphor particle position.
	
\begin{figure}[tb]
		\begin{center}
			\includegraphics[width=6cm,clip]{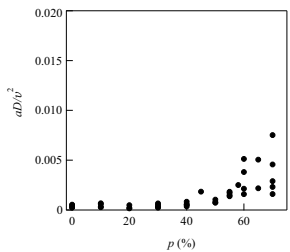}
		\end{center}
		\caption{\label{fig:compare}
			Relationship between $p$ and $aD/v^2$, 
			where $a$, $D$, and $v$ correspond to the sum of sublimation rate and dissolution rate of camphor molecules on an aqueous surface, effective diffusion coefficient, and velocity of a camphor boat, respectively. 
			$aD/v^2$ was much smaller than 1, which suggests our approximation is valid.}
	\end{figure}

	Here, we confirm the acceptability of the assumptions for our experiments.
	We needed values of parameters such as $a$, $D$, and $v$ included in the assumption.
	We used a rectangular camphor boat and chalk powders in measurements of $D$.
	The boat was put on the solution surface covered by the chalk powders, and the camphor diffused into the solution.
	The diffusion was visualized by the chalk powders. 
	We analyzed the videos of powders' motion and estimated $D$.
	A method of the measurement is similar to that in a previous study~\cite{Suematsu2}. 
	The effective diffusion coefficient $D$ against $p$ is shown in Appendix B,
	which shows that $D$ decreases with an increase in $p$.
	For $a$, $a=1.8\times10^{-2}$ s$^{-1}$ was used, which was based on the experimental observation reported in the previous work~\cite{Suematsu2}.
	Using these data, the relationship between $p$ and $aD/{v^2}$ was obtained as shown in Fig.~\ref{fig:compare}. 
	The result shows that the values of $aD/{v^2}$ were sufficiently smaller than 1 for all $p$, 
	which suggests that our assumption is reasonable.
	The result provides the following consideration:
	the camphor concentration around the boat is quite asymmetric, 
	and the decay length of the concentration field at the back of the boat is sufficiently greater than that at the front.
	
	There have been many analytical studies on collective motion of symmetric camphor disks 
	in both experiments and theoretical analyses~\cite{Nishimori,Eric,Ikura,NishiWakai}. 
	There have also been some studies on asymmetric camphor boats, in which numerical calculation  
	for both concentration field and camphor boat positions was performed,  
	and analytical approach under the assumption of slow velocity was  
	performed~\cite{Suematsu,Heisler}. In contrast to these studies, we  
	discussed under the assumption of fast velocity, and this assumption  
	was justified by the experimental observation. It would enable  
	analytical approach on the collective motions of the camphor boats with  
	fast velocity. Therefore, our model would provide a deep understanding  
	of the collective motions on not only camphor boats but also living things.
	
	\section{Conclusion}
	We investigated the velocity $v$ of the asymmetric camphor boat against several glycerol concentration $p$ of the glycerol aqueous solution.
	In order to know the dependence of the camphor boat velocity $v$ on the glycerol concentration $p$, 
	we discussed a numerical model based on a diffusion-reaction equation.
	When it is assumed that the characteristic length of the camphor concentration at the front of the boat is shorter than that at the rear,
	$v$ should obey a power law $v\sim\mu^{-1/2}$,
	where $\mu$ is the viscosity of the base solution.
	The power law agreed with experimental results,
	and it was also confirmed that our assumption in the model was reasonable through a comparison with our experimental results.
	Using our proposed model, we can discuss the profile of camphor concentration, which is difficult to be directly measured in experiments.
	Thus, our experiment has profound significance in the estimation of the concentration through the measurements of the velocity.
	
	As a future topic, it would be worth investigating whether the similar power law $v\sim\mu^{-1/2}$ persists with smaller levels of $v$ in experiments with such variables as increased the boat size.
	In addition, we considered that the hydrodynamic effect was included in the effective diffusion coefficient in this paper.
	It, however, would be also important to consider the fluid flow around the boat
	when we study the behavior of two or more camphor boats as the collective motion.
	As future work, it would be also interesting to consider the hydrodynamic interaction in multi camphor particle system.
	
	\begin{acknowledgments}
		This work was supported by Y. Koyano.
		MS would like to thank Samantha Hawkins of Fukuoka Institute of Technology for proofreading this manuscript.
		This work was supported by JSPS KAKENHI Grant Numbers JP18K11338, JP18K03572, JP25103008 and JP15K05199.
	\end{acknowledgments}
	
	\appendix
	
	\section{Physical properties of a glycerol-water solution as a base solution}
\begin{figure*}[tb]
		\begin{center}
			\includegraphics[width=14cm,clip]{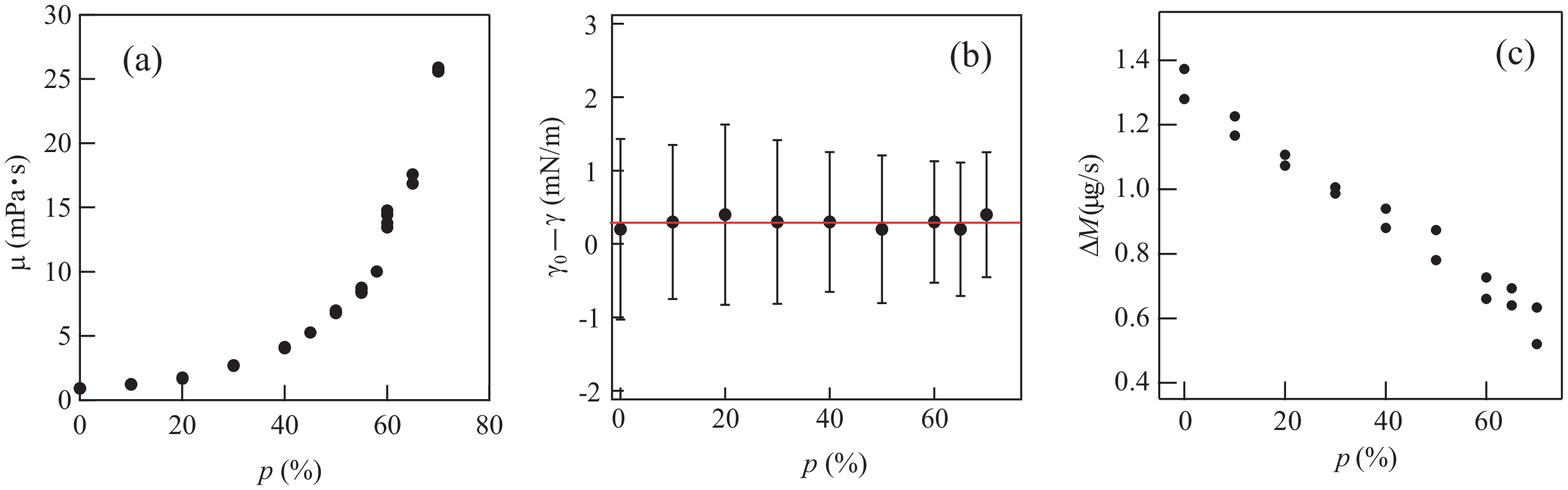}
		\end{center}
		\caption{\label{fig:solution}
			Physical properties of aqueous solutions of glycerol as the base solution.
			(a) Viscosity $\mu$ against glycerol concentration $p$, which is a percentage of glycerol mass in a glycerol-water solution.
			(b) Surface tension difference $\gamma_0-\gamma$ against $p$, 
			where $\gamma_0$ and $\gamma$ are the surface tension of a glycerol-water solution without camphor molecules and that of the solution in which camphor molecules are dissolved, respectively.
			(c) Weight loss $\Delta M$ against $p$.}
	\end{figure*}

	Figure~\ref{fig:solution}(a) shows a viscosity dependence for various glycerol concentrations $p$ of the aqueous solution;
	i.e. $p$ means a percentage of a glycerol mass in the aqueous solution.
	The viscosity $\mu$ was measured using a viscometer (SV-10A, A$\&$D, Japan).
	As shown in Fig.~\ref{fig:solution} (a), the viscosity $\mu$ increased with $p$.
	
	Figure~\ref{fig:solution}(b) shows surface tension difference $\gamma_0-\gamma$ of the solution for $p$,
	where $\gamma_0$ and $\gamma$ correspond to the surface tension for glycerol-water solution without camphor 
	and that for the solution with $6.8 \times 10^{-3}$ g camphor dissolved per 1500 ml, respectively.
	The camphor concentration was set to become close to that in measurements of angular velocity $\omega$.
	The surface tension was measured using a surface tensiometer (DMs-401, Kyowa Interface Science Co., Ltd., Japan).
	The surface tension $\gamma$ with camphor was lower than that of $\gamma_0$ without the camphor,
	and $\gamma$ and $\gamma_0$ decreased with an increase in glycerol concentration $p$.
	The difference of $\gamma-\gamma_0$, however, almost kept constant for different values of $p$ as shown in Fig.~\ref{fig:solution} (b).
	The average value of $\gamma-\gamma_0$ was 0.29 mN/m.
	
	Next, we investigated the dependence of camphor solubility on the glycerol concentration $p$ of the base solution.
	We measured the mass of the camphor disk before and after the camphor disk moved for $\Delta t=50$ min, and the mass change was set to be $\Delta M$.
	From $\Delta M$, we obtained the weight loss rate $\Delta M = \Delta m/\Delta t$.
	As shown in Fig.~\ref{fig:solution} (c), $\Delta M$ decreased with an increase in $p$.

	\section{Effective diffusion coefficient of camphor on solution with several glycerol concentrations}
	The effective diffusion coefficient $D$ of camphor is included in our assumption,
	and the value of $D$ was necessary for checking whether the assumption was reasonable.
	Thus, we measured $D$ for various glycerol concentrations $p$.
	
\begin{figure}[tb]
		\begin{center}
			\includegraphics[width=6cm,clip]{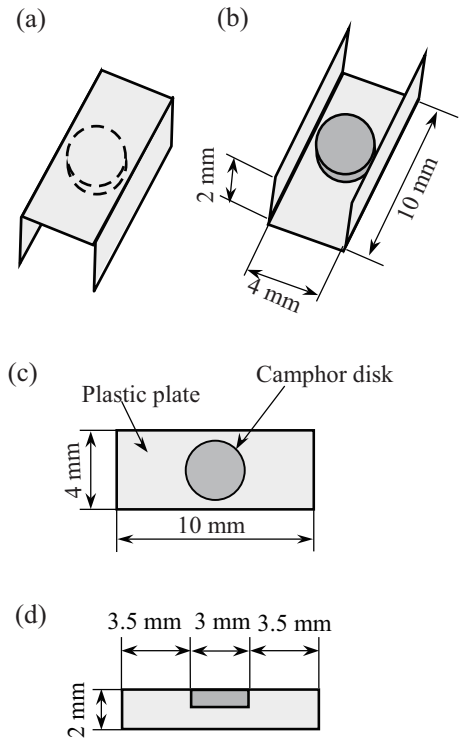}
		\end{center}
		\caption{\label{rectangle}
			Schematic drawings of (a) three-dimensional view, (b) upside down three-dimensional view, (c) top view, and (d) side view of a camphor boat used for the measurements of effective diffusion coefficients.}
	\end{figure}

	The rectangular boat in Figs.~\ref{rectangle}(a)-(d) was used for the measurements of the effective diffusion coefficient of the camphor molecules on the solution,
	and its shape was different from the round-shaped boat in the measurements of the velocity.
	The rectangular boat was made by bending both sides of a rectangular plastic plate that was 8.0 mm in width and 10.0 mm 
	in height at 2.0 mm from the edge. 
	The camphor disk was attached at the center of the plastic plate, where the shortest distance from the edge was 3.5 mm. 
	The shape was similar to the one reported in the previous study~\cite{Suematsu2}.
	
\begin{figure}[tb]
		\begin{center}
			\includegraphics[width=7cm,clip]{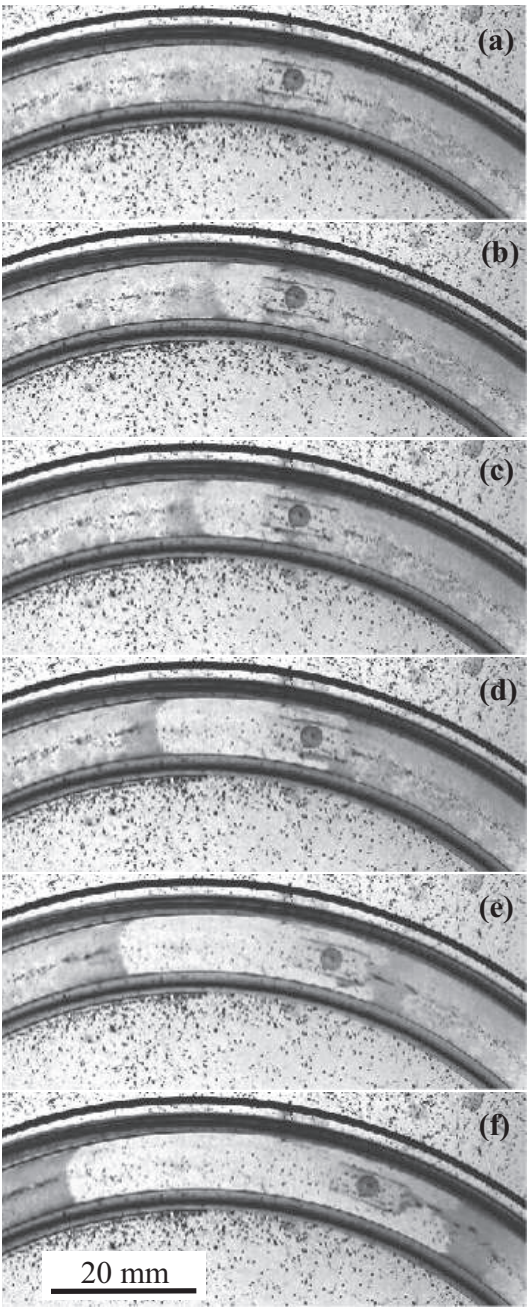}
		\end{center}
		\caption{\label{diffusion}
			Snapshots on the expansion of the camphor molecular layer at (a) $t=0$ s, (b) 0.03 s, (c) 0.07 s, (d) 0.13 s, (e) 0.20 s and (f) 0.30 s, respectively.
			Chalk powders were dispersed on the solution surface for visualization of the camphor layer. 
			The white and gray regions indicate the camphor layer and the region rich in floating chalk powders, respectively.}
	\end{figure}

	Figures~\ref{diffusion}(a)-(f) are snapshots captured from the top at time $t$,
	where (a) $t=0$ s, (b) 0.03 s, (c) 0.07 s, (d) 0.13 s, (e) 0.20 s and (f) 0.30 s, respectively.
	In Fig.~\ref{diffusion}, $t=0$ corresponds to the time at which the chalk powders started moving on the water.
	The diffusion of camphor molecules under the rectangular boat leads to the motion of chalk powders on the water surface.
	As shown in Fig.~\ref{diffusion}(a), all regions of the surface were covered by chalk powders with a gray color at $t=0$.
	The chalk powders started moving at $t=0.03$ s,
	and the water surface without powders was observed as a white region around the boat in Fig.~\ref{diffusion}(b).
	The area of the white region grew with time (Figs.~\ref{diffusion}(b)-(d)).
	The boat stayed at the same position before $t\sim0.2$ s (Figs.~\ref{diffusion} (a)-(e)), although the powders moved.
	The camphor boat, then, started to move after $t\sim0.2$ s (Fig.~\ref{diffusion}(f)).
	In this process, the chalk powders were carried by not only the camphor diffusion but also fluid flow induced by the motion of the boat. 
	
	\begin{figure}[tb]
		\begin{center}
			\includegraphics[width=7cm,clip]{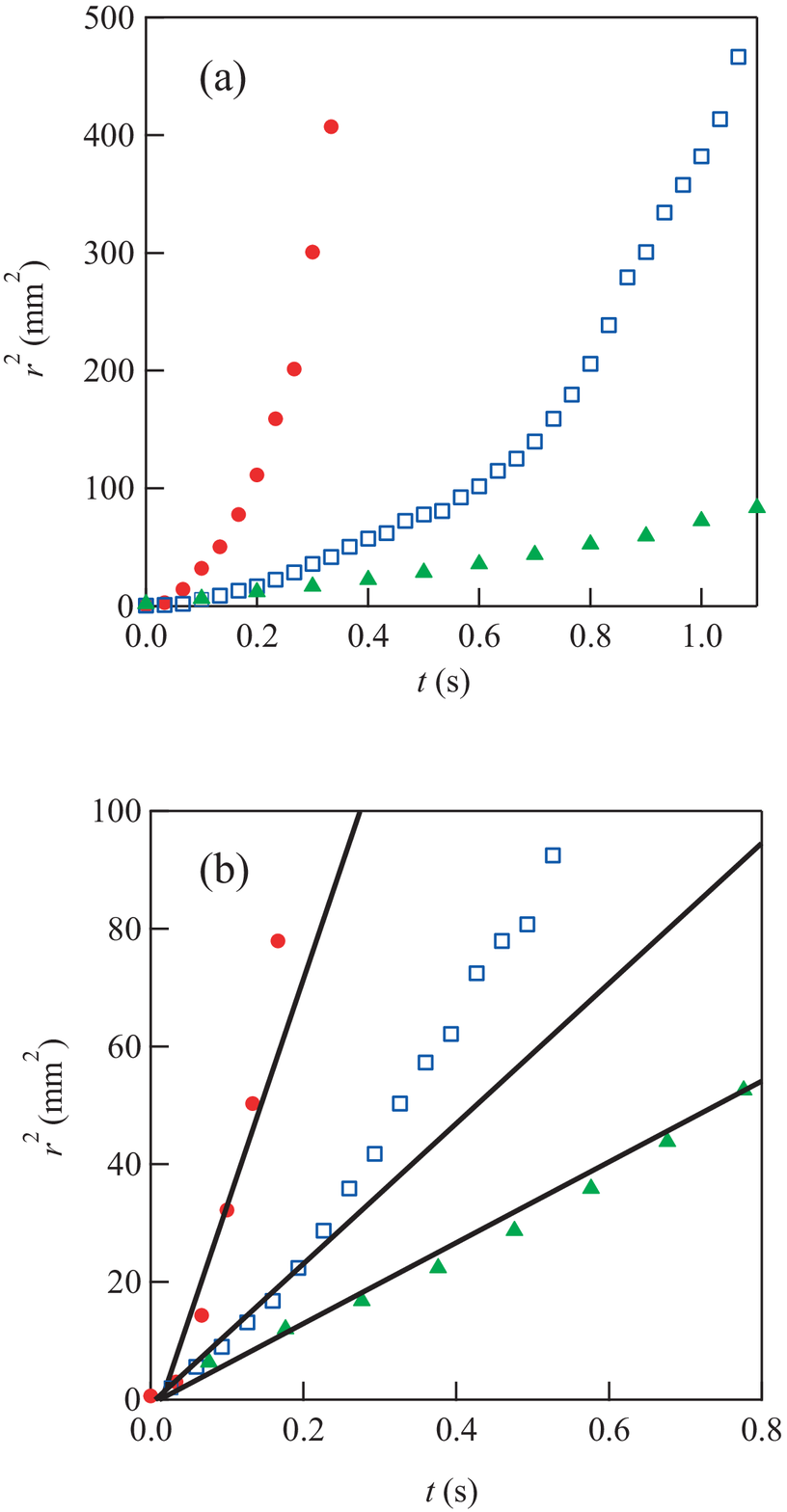}
		\end{center}
		\caption{\label{diffusion2}
			(a) Relationship between time $t$ and $r^{2}$, where $r$ is the longest distance between the edge and the center of the area from which chalk powders were swept out. $t=0$ corresponds to the time at which chalk powders on the solution started moving. 
			Closed circles, open squares, and closed triangles show the data for glycerol concentrations $p=0~\%$ ($\mu=$ 0.92 mPa$\cdot$s), $p=40~\%$ ($\mu=4.03$ mPa$\cdot$s), and $p=70~\%$ ($\mu=25.80$ mPa$\cdot$s), respectively.
			(b) An expanded one for 0 s $<t<0.8$ s in (a), 
			and solid lines are the results of the linear fittings for time before the boat started moving.}
	\end{figure}
	
\begin{figure}[tb]
		\begin{center}
			\includegraphics[width=7cm,clip]{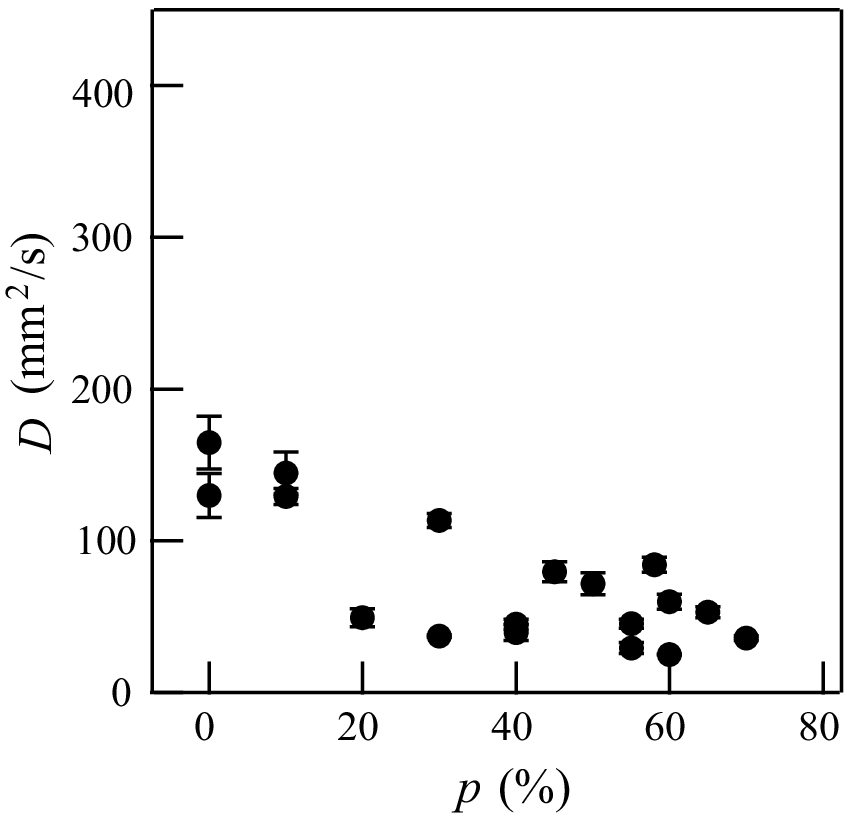}
		\end{center}
		\caption{\label{diffusion3}
			Effective diffusion coefficient $D$ against glycerol concentrations $p$ of the base solutions.
			Error bars denote standard deviations.}
	\end{figure}

	Next, we investigated $r^2$ at time $t$, where $r$ was the longest distance between the edge and the center of the region with the camphor layer, shown as the white region in Fig.~\ref{diffusion}.
	The closed circles, open squares and closed triangles in Fig.~\ref{diffusion2}(a) show the data 
	for $p=0~\%$ ($\mu=$ 0.92 mPa$\cdot$s) for water, and $p=40~\%$ ($\mu=4.03$ mPa$\cdot$s) and $p=70~\%$ ($\mu=25.80$ mPa$\cdot$s) for the glycerol-water solution, respectively. 
	Let us focus on the data for $p=0$.
	The trend of the data changed around at $t\sim0.2$ s, 
	which almost corresponded to the time when the camphor boat began to move as shown in Fig.~\ref{diffusion}.
	As we needed the effective diffusion coefficient of the camphor, 
	we measured $r^2$ in the time range in which the camphor boat did not move.
	Figure~\ref{diffusion2}(b) is an expanded figure for small $t$, i.e. time without the boat motion.
	When the camphor boat stayed at a certain position, $r^2$ increased linearly with time. 
	Linear fittings are shown as solid lines, where fitting was executed for the region 0 s $ <t<$ 0.13 s for closed circles.
	The gradients of these solid lines provide the effective diffusion coefficients $D$ of the camphor molecules on the glycerol-water solution. 
	The effective diffusion coefficient on the water was obtained at $180~(\pm~20)$ mm$^{2}$/s. 
	A previous paper~\cite{Kitahata2} reported that the effective diffusion coefficient $D$ in a numerical study almost agreed with the value for $D$ measured with this method.
	Thus, we consider that this method is reasonable for the measurement of $D$.
	The gradient of the solid line decreases with an increase in the glycerol concentration.
	Figure~\ref{diffusion3} shows the relationship between $p$ and $D$,
	and the tendency that $D$ decreases with an increase in $p$ was confirmed.
	
\begin{thebibliography}{99}
\bibitem{viscous drag} It would be interesting to investigate the viscous dependence for a viscous drag directly in a similar experiment to Ref.~\cite{Suematsu2}.
\bibitem{Delta_M} The experimental data are distributed since the values of $\Delta M$ are noisy as shown in Fig.~\ref{fig:solution}(c) of Appendix A. The values of $\Delta M$, investigated in the motion of the camphor disk, are sufficiently small to be susceptible to external factors.
\end{thebibliography}
\end{document}